# Gauge and metaphysics of spacetime

**Álvaro Mozota Frauca**[1]

**Abstract**
Some authors have argued that spacetime in general relativity should be given a radically different interpretation from the ones given to spacetime in other theories in virtue of it being a gauge theory. In this article I review this sort of argument and argue against this view. That is, I argue that spacetime in general relativity can be understood analogously to spacetime in other models and that the argument from the diffeomorphism invariance of the theory misapplies the concepts of gauge theory to a context in which they are of limited application.

## 1 Introduction

Relativity theory changed our picture of space and time and merged them into a single entity: spacetime. In general relativity our picture of spacetime further changed as spacetime became dynamical. Moreover, general relativity allows for metaphysically interesting phenomena such as singular spacetimes where spacetime itself may begin or end in finite time for certain objects, as well as spacetimes with really peculiar causal structures, including closed timelike curves. On top of this, some authors have argued that the diffeomorphism invariance of the theory is a gauge symmetry and that this fact implies that the departures from the metaphysics of Newtonian space and time are even greater than what I have just discussed. In particular, the influential physicist Carlo Rovelli (2004) has claimed that considering general relativity as a gauge theory has an impact on the observable content of the theory, the philosophers of physics (Earman, 2006; Rickles, 2008) have argued that traditional views of spacetime (relationalism and substantivalism) have to be abandoned and a new, structural-ist one has to be adopted, and Earman (2002) has also argued that traditional views of the metaphysics of time (A and B theories) need to be replaced by a new view.

✉ Álvaro Mozota Frauca
alvaro.mozota@upc.edu

[1] Department of Architectural Technology, Division of Mathematics, Universitat Politècnica de Catalunya, Av. Diagonal 649, Barcelona 08028, Spain

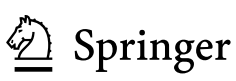

In this paper, I argue against these views and contend that treating the diffeomorphism invariance of general relativity as a gauge symmetry does not force us to add any further modification to our picture of spacetime. In this sense, in this article I am concerned with analyzing and opposing the general argument, which I call the argument from gauge, rather than with discussing those metaphysical positions and interpretations. For this reason, in this article there won't be a detailed discussion of the metaphysics of spacetime, but just a sketch of the different positions available and a series of arguments against the claim that a gauge analysis of the diffeomorphism invariance of general relativity implies changes in the way we should interpret spacetime in this theory that aren't applicable to other spacetimes, such as Newtonian or special-relativistic spacetimes.

I will start in Sect. 2 by briefly introducing general relativistic models and the spacetimes they describe. As I have already mentioned, these structures imply a departure from the space and time of Newtonian models, but I will argue that they are similar to Newtonian space and time in that they consist of sets of points related by causal, geometric, and inertial structures. I will comment on how these models are compatible with different interpretations and philosophical positions such as relationalism, substantivalism, block-universe views of time, or presentism.

Then, in Sect. 3 I introduce the position of the authors who have argued that gauge analysis of general relativity has important metaphysical consequences. I mention the influential work of Carlo Rovelli, although I focus on authors like Earman and Rickles, who make the argument from gauge in a clear and unambiguous way[1]. The arguments by Earman and Rickles make an analogy between the symmetries of general relativity and gauge theories like electromagnetism to argue that the observable content of general relativity is of a particular form, and from there they derive their metaphysical and interpretative consequences.

A key ingredient of the argument I am opposing is that it takes only general relativity and not the other well-known spacetime theories to be susceptible to a gauge analysis and its consequences. However, there are some subtleties involved with this claim, as it is well-known that any spacetime theory can be written in the language of differential geometry and therefore in a way in which one can define diffeomorphism transformations just as in general relativity. I will take this to support my claim that spacetime models are structurally similar and that they ought to be given similar interpretations, but I will discuss how authors like Rovelli, Earman, and Rickles introduce a further distinction here between truly gauge models (sometimes called 'diffeomorphism invariant' or 'background independent') and models which are just expressed in the language of differential geometry (these would be the 'generally covariant' models). In Sect. 4 I will discuss these subtleties and highlight the importance that this difference has for the argument.

[1] There are some passages (Rovelli, 2004, Sect. 2.4.3) in which Rovelli is explicit in that he considers general relativity to be a gauge theory and that this has interpretative consequences, but then he is somewhat ambiguous about what these consequences would be. Then, in his discussions of the metaphysics of Newtonian spacetime (Rovelli, 2004, Sect. 2.4.5) (Rovelli, 2011, Sect. 3) he seems to be endorsing a view in which there wouldn't be a significant difference in the interpretation of spacetime. For this reason, it is unclear to what extent he is endorsing the argument from gauge. For the purposes of this article, it will be enough to address the argument as presented by Earman and Rickles.

Finally, in Sect. 5 I will oppose the argument from gauge. I will adopt three lines of argument. First, I will argue that the argument is too formalistic, in the sense that it follows formal procedures beyond the domain where they are well-established, even if they entail implausible consequences. Second, I will highlight the differences between diffeomorphisms and gauge symmetries in order to show how one should expect that the formal recipes that work for gauge theories do not apply to general relativity. Third, I will put some pressure on the distinction between general relativity and other spacetime theories by introducing a version of electromagnetism that if one followed Rovelli, Rickles, and Earman's arguments shouldn't be interpreted as a gauge theory and therefore should be given a different interpretation. However, I will argue that it describes exactly the same kind of entities, just as general relativity and other spacetime theories describe similar structures. Similarly, I will notice that according to these authors, Minkowski spacetime should receive a different interpretation depending on whether it is considered a model of special or general relativity. I will take that it is an unwanted conclusion of the argument that makes it unattractive. In this sense, I will conclude in Sect. 6 that one should reject the argument from gauge, that is, that one should believe that the differences between how we understand spacetime in general relativity and other spacetimes have to do with the particular geometrical, inertial, and causal structures of spacetime in this theory, but not with the fact that it can be formalized in a way similar to gauge theories.

Before starting, I believe it is important to say a few words about how this debate fits within the existing literature. This debate is clearly influenced and related to the literature on the foundations of quantum gravity and the problem of time that appears at the time of applying canonical quantization techniques to general relativity. In particular, Rovelli is a leading figure in quantum gravity, and his work is very influential and representative of how part of this community thinks. In this sense, the proposals of radical new metaphysics for general relativity by Rovelli, Rickles, and Earman are directly related to the metaphysics of current and possibly future theories of quantum gravity and have an impact on the development of those theories. While the issues surrounding the theory building in the area of quantum gravity are fascinating [2], in this article I will stick to discussing the interpretation of classical general relativity, even if I believe that some of the points I will raise have an impact on the construction and interpretation of quantum theories of gravity. In this context, my arguments can be seen to align with other authors that have opposed a direct application of the concepts of gauge theory to general relativity and its quantization. Let me explicitly mention the work of Pitts (2014, 2017, 2018, 2022) who has defended the standard interpretations of general relativity and has proposed that the notion of observable of gauge theory needs to be modified in order to apply to general relativity. This modification blocks the arguments by Rovelli, Rickles, and Earman, as I will discuss below. Similarly, let me mention the work of Gryb and Thebault (Gryb & Thébaault,

[2] I refer the reader to (Kuchař, 1992) for a discussion of some of the issues that complicate the quantization of general relativity, and to (Wüthrich et al., 2021) and references therein for discussions of the philosophical issues surrounding quantum gravity.

2016, Thébault, 2012, 2021; Gryb and Thébaault 2014, 2023), who have also argued, albeit from a more technical point of view, that diffeomorphisms are different from the gauge transformations of gauge theories, and that they shouldn't be treated and interpreted in the same ways. Nevertheless, I would like to emphasize that although my arguments and position align with these just mentioned, this article offers novel arguments, makes an explicit connection with the debates about background independence and how the interpretations of different spacetime models should be related, and focuses more on conceptual and interpretational issues.

Finally, the work of Gomes is worth citing in this context. The aim of (Gomes, 2022a, b) is to analyze the conceptual similarities and differences between diffeomorphisms and gauge transformations. Interestingly, while the position I am opposing in this article tries to apply the concepts of gauge theory to general relativity, Gomes also makes a unificatory effort but in the opposite direction. That is, Gomes discusses a geometrized version of gauge theories (their formulation in terms of fiber bundles) which makes them similar to general relativity, instead of framing general relativity as a gauge theory. Interestingly, the product of this way of thinking is compatible with the position I am defending in this article. Gomes defends what he calls sophistication, which is the application of the ideas of sophisticated substantivalism also to the case of gauge theory. In this way, while Gomes' approach and interpretation shares the unificatory spirit that makes attractive the idea of applying the concepts of gauge theory to general relativity, his take on how to interpret the theory remains the traditional chronogeometric view of spacetime. While these articles do not address the arguments I aim to rebut in this article, they offer an interesting and complementary analysis that is worth considering.

## 2 Spacetime in general relativity

In general relativity space and time are replaced by spacetime, an entity described by a pair $\langle M, g_{\mu\nu} \rangle$. $M$ is a manifold, i.e., a set of points endowed with certain topological properties such as being 4-dimensional, which represents the set of spacetime points or events. $g_{\mu\nu}$ is the metric tensor defined on this manifold and we can read from it the structure of spacetime. First, $g_{\mu\nu}$ encodes the local causal relations, i.e., whether any pair of points is causally connected or not. This is represented in the lightcone structure that one can derive from the metric and that distinguishes causal curves (timelike and null) from non-causal curves (spacelike). Second, the metric tensor defines a geometry, that is, it defines the lengths (temporal and spatial), areas, volumes, angles, and so on. This geometry influences how physical systems behave. For instance, the readings of clocks[3] are proportional to the proper time defined by $g_{\mu\nu}$ along their spacetime trajectory. Third, $g_{\mu\nu}$ also encodes an inertial structure, that is, it defines the motions that force-free bodies would follow in spacetime.

We can compare the structure of general relativistic spacetimes with the structure of Newtonian space and time. We can represent the spacetime of Newtonian mechan-ics as a triple $\langle M, t_\mu, h_{\mu\nu}\rangle$[4]. Now, instead of having a spatiotemporal metric $g_{\mu\nu}$, we have a temporal one $t_\mu$ and a spatial one $h_{\mu\nu}$[5], which tells us that there is an absolute space and an absolute time. $t_\mu$ and $h_{\mu\nu}$ encode the causal, geometric, and inertial structures of Newtonian spacetime, which are quite different from the structures of general relativistic spacetimes. However, in both cases the interpretation of the models is analogous: in both cases we have a set of spacetime points and we read the causal, geometric, and inertial relations between them from the appropriate tensors. In this sense, the differences between Newtonian space and time and general relativity can be seen as differences in these structures rather than something else arguably deeper.

[3] By clocks one does not need to refer to a human-made system, but one can refer to systems like muons, which have been famously used to test the predictions of relativity. See the discussion in (Fletcher, 2013).

Now we can further compare with the case of special relativity. In special relativity we model spacetime by the pair $\langle M, \eta_{\mu\nu}\rangle$. Spacetime in special relativity is analogous to spacetime in general relativity, the only difference is that we have fixed the spacetime metric to be Minkowski's metric $\eta_{\mu\nu}$. In this sense, the causal, geometric, and inertial structures of special relativity are not that different from what one can have in general relativity, and in particular it is analogous in that it is a spacetime model in which there is no privileged way of splitting into space and time. Moreover, Minkowski spacetime is one valid solution of vacuum general relativity and in this sense it can also be interpreted as a general relativistic spacetime. In this sense, it seems that there shouldn't be any great difference at the time of interpreting special relativistic spacetimes and general relativistic ones.

The difference between special relativity and general relativity is that in general relativity spacetime is dynamical. That is, the properties of spacetime change in different regions according to a set of differential equations, Einstein's equations, which relate them to the matter content of the universe. In this sense, the structure of spacetime is different from model to model and it interacts with matter. This implies a departure from a picture in which spacetime is something that acts but isn't acted upon, which some see as metaphysically problematic. In this sense, general relativity offers this novelty with respect to other spacetime theories, together with a great variety of possible spacetimes that show really interesting features (singularities, for instance) and it is a theory able to explain and predict gravitational phenomena. However, spacetime in general relativity still is a set of spacetime points endowed with a certain causal, geometric, and inertial structure.

This is precisely the position that I want to defend in this article. That is, general relativity is a revolutionary theory that implies that space and time are merged into a single entity and which is dynamical, but its spacetimes are still similar to the spacetimes or spaces and times in other theories as they are a set of events and a series of relations between them. I will argue that considering that general relativity is a gauge theory does not force us to interpret general relativity or general relativistic spacetimes in a way that implies further departures from this position.

[4] I refer the reader to (Kuchař, 1980; Malament, 2012; Knox, 2014; Weatherall, 2021) for discussions of Newtonian space and time from geometrical perspectives similar to the one I am taking here.

[5] These two metrics are orthogonal to each other.

There is a vast literature regarding the best way of interpreting spatiotemporal structures.[6] It is beyond the scope of this article to cover the space of possible ways of understanding spacetime, but I will just mention a couple of distinctions that will be relevant to the discussion here. First, we have the distinction between relationalist and substantivalist views of spacetime. Roughly speaking, for substantivalists spacetime is a real entity that is there independently of matter, while for relationalists what is really real is spatiotemporal relations and not spacetime itself. Second, there are debates concerning the nature of time and whether the past, present, and future are equally real. Metaphysical theories that consider that there is no privileged present are called block-universe views or B-theories, and theories that consider that there is a special now, A-theories.

In the history of physics and of philosophy of physics, the different spacetime models $\langle M, t_\mu, h_{\mu\nu}\rangle$, $\langle M, \eta_{\mu\nu}\rangle$, $\langle M, g_{\mu\nu}\rangle$[7] have been analyzed and used for supporting these philosophical positions. For instance, the isometries of Newtonian spacetime were famously used by Leibniz[8] to argue for relationalism, or the fact that in relativistic spacetimes there is no privileged way of splitting spacetime into space and time has been used to argue for B-theories of time. My position in this article is that the structures of general relativistic models can and should be taken into consideration at the time of interpreting spacetime in this theory, but that the fact that it can be formalized using the structures of gauge theory does not further affect this interpretation.

This is in opposition to the views of Rickles and Earman, who use arguments based on the gauge structure of the theory to argue for their metaphysical claims, as I explain in the next section. Before this, let me advance that there is one issue, the hole argument, in which symmetry transformations play a role, but I will argue that this sort of argument applies to any spacetime theory and that it is a different argument from the one I am concerned with in this article. This reasoning will become clearer after I explain the argument from gauge and will have to wait until the next section.

## 3 The argument from gauge symmetry

The conceptual issues I will be discussing in this paper have to do with the fact that general relativity has a symmetry, diffeomorphism invariance, which complicates its interpretation. The diffeomorphism invariance of general relativity means that for any model $\langle M, g_{\mu\nu}, \phi\rangle$[9] one can build an equivalent one by means of a diffeomorphism. That is, one can build a model $\langle M, g'_{\mu\nu}, \phi'\rangle$ which describes the same set of physical events (spacetime points) and the same causal, geometric, and inertial rela-tions between them. For instance, if a model $\langle M, g_{\mu\nu}, \phi \rangle$ describes the trajectories (in spacetime) of all the bodies of the solar system, a diffeomorphism-transformed version of it, $\langle M, g'_{\mu\nu}, \phi\prime \rangle$, would describe exactly the same trajectories and make the same predictions. That is, we could use either model to predict that some astronomical event will happen, e.g. an eclipse, and when it will happen, i.e., how long an observer, e.g. one on Earth, would have to wait until this event happens. This means that one can choose from an equivalence class of models $\langle M, g_{\mu\nu}, \phi \rangle$ for representing the same spacetime. This is clearly in analogy with gauge theories, where one also has equivalence classes of models that represent the same physical situations. This similarity is interesting and the basis for the positions I want to argue against in this article.

[6] See (Hoefer et al., 2024; Huggett et al., 2024; Savitt et al., 2024) and references therein for detailed discussions on the issues I mention and the metaphysics of spacetime more broadly construed.

[7] Of course, it is a bit anachronistic to write them in the language of differential geometry, but it should be clear that I am referring to the same geometrical structures even at the time they were represented in different forms.

[8] See (Leibniz & Clarke, 1715). Again, the use of the term ‘isometry’ may be anachronistic, but it captures the essence of the argument.

[9] Schematically, $\phi$ represents the matter fields in spacetime.

Consider electromagnetism. It can be formulated entirely in terms of the physical electromagnetic field $F_{\mu\nu}$ which affects the motion of charged particles according to the Lorentz force and which evolves satisfying Maxwell's equations. Alternatively, one can formulate the theory in terms of the 4-potential $A_\mu$, which is convenient, as it allows defining electromagnetic energy and expressing the theory in the Lagrangian and Hamiltonian formulations[10]. However, the 4-potential representation of the electromagnetic field is not unique, which means that there is a whole equivalence class of 4-potentials $A_\mu$ which represent the same physical electromagnetic field $F_{\mu\nu}$.

This is the paradigmatic example of a gauge theory. In gauge theories we have equivalence classes of different representations of the same physical situation. In the context of a gauge theory it is important to distinguish the physical content of a model from just the mathematical details of this representation which do not represent anything physical. In the case of electromagnetism, this means that details of the exact value of $A_\mu$ at some point are not physical and change from representation to representation, while the value of $F_{\mu\nu}$ at that point is entirely physical and all the representations agree on it[11]. Similarly, in general relativity, the numerical values of the metric tensor's components at a spacetime point depend on the chosen coordinate system, varying across different representations. The time that an observer experiences between two events is independent of the mathematical details of the representation chosen for describing spacetime, and is part of the physical content of the general relativistic model.

More generally, we can say that all the diffeomorphism-related models of general relativity describe the same set of spacetime points and the same spatiotemporal relations between them. In this sense, we have a perfectly fine understanding of the theory and we are able to distinguish the physical content of the theory from the mathematical details of the model with no physical correlate. However, this intuitive account of general relativistic models and the way we interpret them is not satisfactory for those with more formal intuitions.

[10] Here I am concerned just with classical theories, and hence I won't be concerned with the role that gauge quantities could play in quantum versions of these theories. In the case of electromagnetism, some have argued that the Aharonov-Bohm effect means that the 4-potential $A_\mu$ is physical. In this article I will leave those considerations aside and discuss only the classical version of the theory. I refer the interested reader to (Belot, 1998) for a discussion of this point.

[11] The arguments by Rovelli, Earman and Rickles are based on this understanding of gauge theory and electromagnetism, and for the goals of this article this characterization will suffice. However, let me notice that there are more nuanced analyses of gauge theories, gauge transformations, and the physicality of gauge fields. I refer the reader to (Teh, 2016; Weatherall, 2016) and references therein for some such nuanced analyses.

The reason for this is that in gauge theories like electromagnetism one can give an entirely formal definition of the quantities that directly represent the physical content of a model[12]. Gauge transformations in the Hamiltonian formalism are generated by some particular functions $G$[13], and one can define the gauge invariant content of the theory as the quantities that do not change under these transformations. More precisely, one defines gauge invariant quantities $f$, also known as observables, to be the ones that satisfy:

$$\{f, G\} = 0\,, \tag{1}$$

where the brackets represent the Poisson brackets of the phase space of the theory. In the case of electromagnetism one can show that $F_{\mu\nu}$ satisfies this condition while $A_\mu$ does not. For this reason, it has been argued that the physical content of any gauge theory lies in the observables that satisfy this condition[14].

This formal requirement works fine for theories like electromagnetism, but it is problematic in the case of general relativity. Quantities like 'the time along a time-like curve' which are part of general relativistic models are hardly, if possible at all, expressible as phase space functions that satisfy condition 1. If we apply the criterion that the physical observables need to satisfy this condition, we are in a position in which most of what we took to be part of the physical content of general relativity is not, or at least not in a straightforward way. If we insist on this kind of reasoning we may reach quite surprising conclusions, which I will argue against in this article.

An author who has had a clear influence with respect to this kind of argument is Carlo Rovelli[15]. He has taken criterion 1 to imply that the physical content of general relativity shouldn't be expressed in terms of causal, geometrical, and inertial structures as I have discussed, but in terms of observables, i.e., phase space functions which satisfy 1. These observables would have a strong relational character and would correspond to quantities like 'the value that a field $\varphi$ takes when other four fields $\phi^\mu$ take certain values.' There is some ambiguity in Rovelli's work regarding

[12] Changes in these quantities necessarily correlate with changes in the world.

[13] See (Pooley & Wallace, 2022; Pitts, 2024; Mozota Frauca, 2024) for a recent debate about the form of these functions.

[14] The history of this sort of claim in the context of general relativity goes back to the work by Bergmann in the sixties (Bergmann, 1961). Another influential author in this area was Dirac, who developed his quantization procedure making use of the constrained formalism for dealing with gauge symmetries (Dirac, 1964).

In more complicated gauge theories like Yang-Mills theories, observables are not the values of fields at a spacetime point, but line integrals of fields. Quantities like the forces that Yang-Mills fields exert on bodies carrying Yang-Mills charge are also invariant.

[15] He has developed his position regarding observables in several publications during his career. The most relevant ones are (Rovelli, 1991, 2002a, 2002b, 2004).

which implications he takes this to have for the nature of spacetime[16], but he certainly endorses the view that the physical content of general relativity needs to be represented by this sort of phase space function and his view has certainly influenced the other authors I will discuss in this article.

Philosophers of physics like Earman and Rickles have followed Rovelli's line of argument and concluded that if one takes the content of general relativity to be encoded in the observables, then one is forced to take a (ontic) structuralist interpretation of the theory[17]. That is, while one can interpret spacetime in Newtonian theories or in special relativity in either a substantivalist or a relationalist fashion, these metaphysical positions wouldn't be allowed in general relativity according to these authors. The reason for this would be that when one considers observables (in the sense of equation 1) one moves away from the standard manifold picture.

Similarly, Earman takes this criterion for defining observables to have implications for the metaphysics of time[18]. While in standard spacetime theories one has a debate between presentism and block-universe views of time, these alternatives would be inferior to a new metaphysics of time that would be motivated again from the form of observables in general relativity, which are supposedly independent of manifolds and similar structures.

In both cases we find similar situations: we have a metaphysical debate about how to understand and interpret the spatiotemporal structures that appear in our physical models, that is, substantivalism vs relationalism and presentism vs block-universe views of time, and these authors take the gauge structure of general relativity to imply that in both debates one should take a third, more radical, and new perspective. In this article I do not want to enter into the details of any of those debates, but I will just argue that the argument from gauge is not appealing at all, as it relies heavily on a formal criterion that can be challenged, as I will do in Sect. 5.

The argument from gauge usually comes accompanied by some version of the hole argument, and it will be helpful to make a brief comment about it and how it is related with the general argument in this article. It is beyond the scope of this article to add to the vast literature on the topic[19], and I will just focus here on the way Rovelli, Earman, and Rickles use or mention this argument to support their views. The hole argument exploits the diffeomorphism invariance of general relativity to argue that a naive version of substantivalism would have to accept a sort of indeterminism. The reason for this is that given the diffeomorphism invariance of the theory, given a set of initial conditions, the set of spacetime points or events and the spatiotemporal relationships between them are determined, but not which point in the manifold will correspond to each spacetime event. In other words, a set of initial conditions predicts that an event $E$, say the collision of two stars, will occur, but it does not single out if it will be represented in the model by a point $P$ in the manifold or by another one $P'$ . If one reads the model too literally and considers that these are two different possibilities, then one faces indeterminism.

[16] For instance, in his book (Rovelli, 2004) there are some passages in which he takes the nature of coordinates to be radically different in general relativity, but then there are some other passages in which he analyzes Newtonian physics in a way which is very analogous to the way he analyzes general relativity. In any case, in his book he does not go as far in his claims as the authors cited below do.

[17] This is argued for in (Earman, 2006; Rickles, 2008).

[18] See (Earman, 2002).

[19] Some of the most relevant pieces on the topic are (Earman & Norton, 1987; Hoefer, 1996; Weatherall, 2018, 2020; Norton & Zalta, 2019; Roberts & Weatherall, 2020; Pooley, 2022).

However, virtually no one holds this sort of naive reading of general relativistic models, and the most widespread view is that one shouldn't take individual models $\langle M, g_{\mu\nu}, \phi \rangle$ but equivalence classes of them to constitute the starting point for the interpretation of the theory. Point $P$ in one model, and $P'$ in the other are just different representations of one physical event $E$, and nothing changes, physically or metaphysically, if we choose one model or the other to represent spacetime. From here, two standard positions or families of positions arise: substantivalist positions (sometimes called sophisticated substantivalist to distinguish them from the naive view) and relationalist ones[20]. That is, after considering equivalence classes of models it is still possible to argue that spacetime is fundamental or that spatiotemporal relationships between matter are.

I take that this is in a nutshell the standard analysis of the hole argument in the literature. Let me briefly turn now to the discussions of the argument by Earman and Rovelli. In (Earman, 2002), Earman uses a hole-like argument to support the argument from gauge regarding the metaphysics of time. Here I won't discuss his argument in detail, but just note that the standard views in the metaphysics of time accommodate the fact that one has to consider equivalence classes of models under diffeomorphism without falling into Earman's position and I refer the reader to (Maudlin, 2002) for a rebuttal of this argument. In (Rovelli, 2004, Sect. 2.4.2), Rovelli makes an interesting observation about the hole argument: he argues that it gets substantivalist and relationalist positions closer, maybe to an extent that the distinction stops making sense. This idea fits nicely with Earman's proposal of a new structuralist metaphysics that leaves behind substantivalism and relationalism. Again, it is not the purpose of this article to evaluate this sort of argument, but just to insist that the standard view in the metaphysics of spacetime is that substantivalism and relationalism are two different positions compatible with the hole argument.

What is more relevant to this article is to notice that the hole argument relies on the diffeomorphism invariance of general relativity on a purely kinematical level. That is, it is just based on the fact that diffeomorphism-related models are considered equivalent and the fact that spacetime is dynamical does not play a role in it. This observation allows us to formulate a version of the hole argument for Newtonian spacetime or Minkowski spacetime in special relativity. Just as in the case of general relativity, given a set of initial conditions, it is not determined which manifold point $P$ will correspond to each event $E$ in the formulation of these models in the language of differential geometry ($\langle M, t_{\mu}, h_{\mu\nu} \rangle$ or $\langle M, \eta_{\mu\nu} \rangle$). This observation would work to rule out[21] naive interpretations of these models, but leaves room for the standard relationalist and substantivalist views of spacetime in these theories.

In this sense, even if the hole argument was originally formulated in the context of general relativity, it can be formulated for other theories. As it is an argument based just on the symmetry at this kinematical level, i.e., it works independently of whether the symmetry is dynamical or not, it does not pick up general relativity as a special case. Therefore, it cannot support the conclusion of the argument from gauge that spacetime in general relativity needs to be interpreted in a different way.

[20] I refer the reader to (Hoefer, 1996; Norton & Zalta, 2019; Roberts & Weatherall, 2020) and references therein for pieces in which these standard positions are outlined.

[21] Namely, to make them indeterministic and unattractive.

In other words, my analysis of the hole argument is that it is an issue that has to do with the interpretation of the mathematical structure of our spacetime models. As I argued in Sect. 2 the formal analogy between different spacetime models invites us to interpret them in similar ways. What is important to notice is that while the hole argument applies to any diffeomorphism invariant model, the argument from gauge is supposed to apply only to theories for which diffeomorphism invariance is a gauge symmetry. At this point, we need to clarify what different authors mean by 'gauge symmetry' and whether it applies to the spacetime theories we are considering.

## 4 Is diffeomorphism invariance a gauge symmetry? Dynamical and kinematical meanings of 'gauge symmetry'

As we have seen, the argument above relies on taking the diffeomorphism invariance of general relativity to be a gauge symmetry in order to be able to apply the criteria to identify the physical content of gauge theories. However, diffeomorphism invariance is not exclusive to general relativity.

Both Newtonian models $\langle M, t_\mu, h_{\mu\nu}, \phi\rangle$ and special relativistic models $\langle M, \eta_{\mu\nu}, \phi\rangle$ are diffeomorphism invariant in the sense that one can build an equivalent model by transforming the metric tensors $t_\mu, h_{\mu\nu}$ or $\eta_{\mu\nu}$ and the matter fields $\phi$. This is just the active version of the passive transformation corresponding to a change of coordinates. Any spacetime model can be made diffeomorphism-invariant or generally covariant. In this case, we have equivalence classes of models just as we had in general relativity and gauge theories. This invites us to consider that all spacetime models are to be interpreted in similar ways and that they all are gauge theories to the same extent.

Authors like Earman and Rovelli have argued against this line of argument[22]. They argue that there is a difference between a diffeomorphism invariance like the one in Newtonian spacetime and special relativity and the diffeomorphism invariance of general relativity. Indeed, they introduce a distinction between 'generally covariant' (in Rovelli's terms) or 'formally generally covariant' (in Earman's) and 'diffeomorphism invariant' (Rovelli) or 'substantively generally covariant' (Earman) theories that aims to grasp this difference. For this, we need to introduce a distinction between two kinds of variables in our theories: dynamical and fixed or background variables.

Intuitively, a fixed variable represents a structure or entity that is the same in every physical model of a theory. For instance, space and time in Newtonian mechanics would be a fixed variable, given that it is the same for every possible model. On the other hand, the matter and field content of spacetime would be represented by dynamical variables, given that in different models one can have different matter configurations behaving differently. From a formal perspective[23], when one defines models in the Lagrangian formalism, dynamical variables are variables that are varied in an action principle and have associated equations of motion, while background variables are not variables that are to be varied in an action principle and that many times do not even explicitly appear as variables in our formalization of the models. Following this definition, spacetime is a background or fixed structure in Newtonian mechanics and special relativity and a dynamical one in general relativity[24].

[22] See (Rovelli, 2004; Earman, 2006).

Now, Rovelli restricts the use of the term ‘diffeomorphism’ only to the case when it is a transformation that affects only dynamical variables. In the case of Newtonian and special relativistic models, this would exclude diffeomorphisms that affect the non-dynamical $t_\mu, h_{\mu\nu}$ or $\eta_{\mu\nu}$, and hence, these theories wouldn’t be diffeomorphism invariant for Rovelli, even if they would still be generally covariant. Similarly, for Earman, only in the case of general relativity we would have a ‘substantive general covariance’. In this article I will stick to the terminology I have been using, namely, by ‘diffeomorphism’ I will be referring to any diffeomorphism, not just to those affecting dynamical variables[25]. Besides the terminological difference, for these authors, this difference is relevant in this context because it entails a difference in the way the dynamics is formalized. It is only in the case of general relativity that one has to use a Lagrangian with a dynamical symmetry (this is a symmetry that affects dynamical variables), just as in the case of gauge theories like electromagnetism. This motivates applying the definition 1 to the case of general relativity, which is the basis of Rovelli, Earman, and Rickles’s arguments. In the case of theories with non-dynamical spacetimes, there is no dynamical symmetry related to diffeomorphisms, and the analogy with gauge theories cannot be established at the level of the way the dynamics is formalized[26].

In this sense, we can see that we have two different notions of gauge symmetry: a kinematical and a dynamical one. We say that a theory has a kinematical gauge symmetry if by applying some transformation to a given model one can build an equivalent one. A theory has a dynamical gauge symmetry if it has a kinematical gauge symmetry and, on top of this, the symmetry transformation affects only dynamical variables. When expressed in the Lagrangian formalism, dynamical symmetries appear as transformations of the action that leave the action invariant and that only affect dynamical variables. Authors like Earman and Rovelli would only call this kind of symmetry ‘gauge symmetries’, but it is important to notice that both senses can be useful in given contexts and that using one or the other can make a difference in this discussion.

[23] See (Pooley, 2010, 2017) for discussions of subtleties related with these definitions.

[24] Here I am following the standard presentation, although there are some authors who have argued that depending on which action principle one uses for encoding general relativity, one can reach different conclusions about whether the determinant of the matrix components of metric is dynamical or not. In that case, one should rather say that most of the spacetime metric is dynamical. See (Pitts, 2006).

[25] See (Rovelli, 2004; Earman, 2006) for discussions of their terminology and (Pooley, 2017) for a careful and complete discussion of this issue. Notice that in the literature the terms ‘diffeomorphism invariance’ and ‘general covariance’ have been used with different meanings, and some authors have used ‘general covariance’ for referring to the supposedly physically meaningful one.

[26] I refer the reader interested in the technical results behind these claims to (Pons et al., 2010; Pitts, 2014; Mozota Frauca, 2023) and references therein.

When we analyze diffeomorphism invariance as a kinematical gauge symmetry, what we find is that there is no important difference between different spacetime theories. At a kinematical level, diffeomorphism invariance is not exclusive of general relativity, and the interpretation of spacetime models should be quite independent of the theory: we have a model or an equivalence class of models out of which we read the causal, geometric, and inertial structure of spacetime. Just from analyzing models at this kinematical level it seems that one would have never reached the conclusion that one should interpret general relativistic models differently from other spacetime models.

It is therefore key to emphasize that it is the dynamical notion of gauge symmetry that would be making a difference for authors like Earman, Rovelli and Rickles. The condition 1 arises in the Hamiltonian formalism only for dynamical variables, given that one does not treat fixed variables as variables to be varied, and only dynamical symmetries explicitly appear in the formalism. As the argument from gauge starts from this condition to reach its conclusions, it is an argument that applies only to dynamical symmetries, and for this reason it would be making a difference between the different spacetime models depending on whether they describe dynamical or non-dynamical spacetimes, even if they share the same kinematical gauge symmetry. This is the reason why Rovelli and Earman wanted to distinguish between different kinds of diffeomorphisms.

One of the reasons I will give in the next section for opposing this argument is precisely that from my point of view it is not sensible to change our interpretation of a model with a kinematical gauge symmetry depending on whether we consider it as pertaining to a theory in which this symmetry is dynamical or not.

## 5 Rebutting the argument from gauge

Having introduced my position and the argument from gauge and having clarified the particular role that dynamical symmetries play in the argument, in this section I will develop three objections to the argument from gauge.

### 5.1 Too formalistic an argument

The first argument I present is that Rovelli, Earman, and Rickles' reasoning is excessively formalistic, relying too strictly on formal procedures to derive their interpretations. They use the Hamiltonian formalism to derive claims that are to a large extent different from any claim that had been made by analyzing the theory in its standard or Lagrangian forms. While some authors have taken this surprising feature to be indicative that something has gone wrong with the analysis in the Hamiltonian for-malism[27], Rovelli, Earman, and Rickles stick to their conclusions no matter how radical or incompatible with other accounts of general relativity they are. In this sense, the question we should ask ourselves is, what should we trust more, the formal recipes that work for gauge theories like electromagnetism or the best interpretations of general relativity that we had before applying those recipes?

My answer is that we should rather take the interpretations of general relativity that were developed and established for the theory. Arguments relying on the formal structures of our theories may work and may be helpful for clarifying the content of a theory, but one should be careful not to take them too far. One can distinguish (at least) two parts in a physical theory: a mathematical formalism and an interpretation linking mathematical variables and structures with entities in the world described by that theory. While studying the formalism can help us answer questions about those entities in the world, it should be clear that the dependence relation goes in the opposite direction: the formalism depends on the entities in the world and not the other way around.

In the case of electromagnetism, we started with a clear picture of what the physical content of the theory was: an electromagnetic field that mediates the interactions between charged particles and tells them how to move. From this, we can derive that if we use the 4-potential Hamiltonian formalism to formalize this theory, it is the quantities defined by 1 that represent the electromagnetic field and the physical content of the theory. But it should be clear that we start from a clear interpretation of the theory to arrive at the formal result.

In the case of general relativity, we also have a perfectly fine interpretation of the theory (or some equally reasonable ones) which tells us that $\langle M, g_{\mu\nu} \rangle$ describe a set of spacetime points together with a causal, geometric, and inertial structure. If now we try to force upon the formalism of general relativity the recipes that worked in the case of electromagnetism and they fail to give us our interpretation of the theory, that's too bad, but as we take our theory and its interpretation to be prior to the formalism, this just means that the formal recipes that work for one theory do not work for the other.

### 5.2 Disanalogies between GR and gauge theories

One could just stop at this point, but to have a more complete story one can analyze why recipe 1 works for finding the physical content of theories like electromagnetism while it fails for theories like general relativity. The reason for this is that the symmetry transformations in both cases are fundamentally different. More precisely, while gauge transformations in electromagnetism can be understood in both a local and a global way, diffeomorphism invariance can be understood as a gauge symmetry only in a global way. This observation is not new, and it has been raised from different perspectives in (Kuchař, 1993; Pons et al., 2010; Pitts, 2014, 2017, 2018; Mozota

[27] See (Kuchař, 1993; Pons et al., 2010; Pitts, 2014, 2017, 2018; Mozota Frauca, 2023, 2025, 2024, 2026; Thébault, 2021; Gryb & Thébaault, 2023, 2016, 2014). Particularly telling is the subtitle in the rebuttal by Maudlin (2012) of Earman's arguments: 'How to Abuse Gauge Freedom to Create Metaphysical Monstrosities'.

Frauca, 2023, 2024; Thébault, 2021; Gryb & Thébault, 2016, 2023; Pitts, 2022). Here I will briefly discuss the disanalogy at a conceptual level, and I refer the reader to the references for more detailed, technical, and historical discussions.

Both diffeomorphisms in general relativity and gauge transformations in theories such as electromagnetism can be understood as gauge symmetries from a global point of view: they map solutions of the equations of motion into physically equivalent solutions of the equations of motion. Both define equivalence classes of mathematical representations of the same spacetime or of the same history of fields like the electromagnetic one. However, they differ in the way they act locally.

By the local action of a gauge transformation, I mean the effect that the transformation has at a spacetime point or instant of time. That is, instead of looking at complete solutions of the equations of motion representing evolutions in spacetime we will consider what happens at a coordinate point $x^\mu$ and what is represented by this point. Notice that the term 'local' appears in the context of gauge theories also with other meanings[28], and hence it is important to make clear the sense in which I am using it.

Consider the action of an electromagnetic gauge transformation at a spacetime point $x^\mu$. It transforms a 4-potential $A_\mu(x^\mu)$ into an equivalent one $A'_\mu(x^\mu)$, while it leaves the electromagnetic field at that point untouched ($F'_{\mu\nu}(x^\mu) = F_{\mu\nu}(x^\mu)$). In this sense, for this theory it makes complete sense to claim that what is observable at the spacetime point $x^\mu$ is the value of the electromagnetic field $F_{\mu\nu}(x^\mu)$ and one can argue that the local version of the transformation is a gauge transformation: it is a transformation that affects the way we represent physical facts ($A_\mu(x^\mu)$) at one point but not these physical facts. For this reason, one can claim that the observables at the point $x^\mu$ are the quantities that remain invariant under such a transformation, which is equivalent to definition 1.

Now consider a general relativistic model in which we could have some matter fields. For the sake of the argument assume that one field $T$ represents the temperature at each spacetime point. The action of a diffeomorphism from a local perspective is to map every point in the manifold to a different one. This means that at a point $x^\mu$ the transformed temperature field will tell us the temperature that was originally assigned to a point $x'^\mu$. Now it should be pretty obvious that one cannot apply the same logic as in electromagnetism. Now the claim that the physical content of the theory at point $x^\mu$ is what remains invariant under the diffeomorphism does not make sense, as the point $x^\mu$ now represents a different spacetime point[29]. In this sense, the value of temperature field at that point $T(x^\mu)$ and the value of its transformed version $T'(x^\mu)$ do not represent the same state of affairs at a point in different fashions, but they represent the temperatures at two different spacetime points. To make the case more vivid, $x^\mu$ before the diffeomorphism could represent the spacetime event in which I am typing these words, while after the diffeomorphism it could represent a point just a few instants after the Big Bang. Clearly, the temperature at those two points is quite different, and it does not make sense to look for the invariant content at $x^\mu$.

[28] There are at least two other meanings with which the term 'local' is used. One speaks about local symmetries to refer to symmetry transformations that vary arbitrarily on the point it is applied. Diffeomorphisms are clearly local in this sense, which is not the one I am intending here. The term local observable is also used to refer to some mathematical properties of power expansions, and it won't be relevant here.

[29] Notice that in this discussion based on the Hamiltonian formalism I am describing diffeomorphisms as a transformation $T(x^\mu) \to T'(x^\mu)$. This transformation can be understood in an active or passive way depending on whether one considers that one is 'moving' the field $T$ or just changing the coordinates. This difference plays a role in the hole argument (see again Sect. 3 and the references cited there), but for the argument here it is not important. That is, as we are rejecting interpreting bare points of the manifold directly as spacetime points, we reach the conclusion that after applying the transformation $T(x^\mu) \to T'(x^\mu)$ by $x^\mu$ we are referring to a different spacetime point, independently of whether we represent this point with a different point in the manifold or the same point as before the transformation.

This is, of course, extendable to any general relativistic model. It does not make sense to look for the invariant content at a coordinate point $x^\mu$ or to understand the gauge transformation in a local sense, because the meaning of $x^\mu$ changes with the transformation. As a consequence, there is no motivation for defining observables using the condition 1. The temperature field is very reasonably part of the physical content of the model and 'observable' in the most intuitive sense, despite not satisfying condition 1.

Summing up, one can distinguish two types of gauge transformations. On the one hand, we have transformations like electromagnetism which do not 'move things in spacetime' and which can be understood in a local way. For this kind of transformation one can apply the recipe 1. On the other hand, we have transformations that change which spacetime point is assigned to each point in our representation. In this case one cannot understand the transformation as leaving the physical properties at a representation point intact, and the recipe does not apply. Because diffeomorphism invariance falls into this latter category, this blocks the arguments for the authors I am opposing in this article.

### 5.3 Interpretation of non-dynamical and dynamical entities

The last argument I want to present focuses on the fact that the arguments I am opposing rely heavily on the fact that spacetime is dynamical in general relativity in order to claim that its diffeomorphism invariance is a dynamical gauge symmetry and not just a kinematical one and that this has the deep metaphysical implications that it is supposed to have. Here I want to challenge the claim that a gauge symmetry deserves a different interpretation if it is a dynamical symmetry by considering first what would happen if we had a fixed electromagnetic field, with just a kinematical gauge symmetry instead of a dynamical one. My position is that the interpretation would be the same, and hence, that the interpretation of spacetime structures shouldn't change depending on whether they are dynamical or not. Indeed, I will also consider the interpretation of Minkowski spacetime as a model of general relativity and as a model of special relativity. I will argue that it should be interpreted in similar ways in both, and that the position of Rickles and Earman that the same mathematical object represents two radically different metaphysical possibilities seems to be begging the question.

Consider a situation in which we have a set of charged particles moving under the influence of a strong external electric field, such that the interactions between those particles can be neglected for all practical purposes. Alternatively, one can consider a world in which charged particles feel the electric field but they do not act as sources of it. In both cases, we would write a theory describing the behavior of the particles which would amount to the Lorentz force for a fixed, non-dynamical $F_{\mu\nu}$. If we expressed this in terms of the 4-potential[30] we would find a kinematical gauge symmetry: we would have an equivalence class of 4-potentials $A_\mu$ which work equally fine for describing the fixed electric field. In this case, as the field is not dynamical we could say that we have a gauge symmetry just in the kinematical sense. That is, as the symmetry does not concern dynamical fields, we won't have singular Lagrangians and we won't have to use the constrained Hamiltonian formalism to deal with this symmetry. Now the relevant question is, should we also change in some way the interpretation we give to $A_\mu$ or $F_{\mu\nu}$?

My answer is that we should not. The electromagnetic field is still the entity that tells charged particles how to move, despite being fixed in this case. The physical content of the theory is essentially the same, once we accommodate the fact that the electromagnetic field is non-dynamical. In this sense, the observable content of the model is the same in the dynamical and non-dynamical cases: an electromagnetic field and the positions of the particles. This is so even if only in the dynamical case one can appeal to the definition 1 in the Hamiltonian formalism. As far as I can see, there is no good reason to claim that there should be any difference in the way we interpret the nature of the electromagnetic field depending on whether it is dynamical or not.

The same conclusion should hold in the case of spacetime. A particularly surprising consequence of the argument from gauge is that if we follow it we have to make a different interpretation of Minkowski spacetime depending on whether we understand it as a model of special relativity or a particular solution of general relativity. If one takes it to be the fixed spacetime of special relativity, then one can interpret it in either a substantivalist or relationalist way and one can argue for either a presentist or a block-universe view of time in it. Similarly, there is no problem in getting to know the physical content of such a model, and a quantity like the value of a field at a point is considered an observable. Now, if we consider this spacetime to be a model of general relativity, our interpretation of it should be radically different according to the argument from gauge.

From my point of view, this is clearly an unwanted consequence of the argument. For Earman, however, this seems to be an acceptable consequence of the argument, as he discusses (Earman, 2006, p. 13) that the argument from gauge distinguishes general relativity from other spacetime theories and he explicitly includes vacuum solutions of general relativity. Rickles, on the other hand, is not so clear on this point. In his discussion of the argument from gauge (Rickles, 2008, Sect. 4), Rickles relies on the gauge structure of general relativity to argue for structuralism in a way that does not distinguish vacuum solutions from non-vacuum solutions. In this sense, we can see how the argument from gauge for him makes a difference between general relativity and theories like special relativity. However, in the same article, when he discusses a different sort of argument related to background independence (Rickles, 2008, Sect. 3), he claims that the metric in vacuum general relativity is an absolute object, which implies that the theory wouldn't be interpreted in the same way as general relativity with matter. In this sense, one could read Rickles as implying that the argument from gauge only works when matter is present. But the gauge structure of general relativity is the same independently of whether matter is present or not, and hence one would need a refined version of the argument from gauge to avoid the consequence that Minkowski spacetime would be interpreted differently depending on whether it is considered a general relativistic model or a special relativistic one.

[30] A reason for this could be to write a Lagrangian for the particles.

For these reasons, I believe that there is no compelling justification for claiming that the structures in our physical models should be given radically different interpretations depending on whether they are dynamical or not. This applies both to the case of fields like the electromagnetic field and the spacetime structures of our physical models. From my point of view, this is clearly an unwanted conclusion from the argument from gauge that makes it less attractive.

## 6 Conclusions

A spacetime model describes a set of spacetime points, the relations between them, and what bodies, matter, and fields do in this spacetime. I have argued that all spacetime models can be formulated in diffeomorphism invariant ways and that the fact that in general relativity this invariance is a dynamical symmetry does not affect its interpretation. In this sense, I have argued that the argument from gauge should be rejected.

**Acknowledgements** I want to thank Guy Hetzroni and the audience at the 9th Meeting of the Society for the Metaphysics of Science for their helpful comments on an early version of this article. I am also grateful to the reviewers of this journal for their suggestions.

**Author contributions** NA.

**Funding** Open Access funding provided thanks to the CRUE-CSIC agreement with Springer Nature.

**Data availability** NA.

**Materials availability** NA.

**Code availability** NA.

## Declarations

**Ethical approval** NA.

**Consent to participate** NA.

**Consent for publication** NA.

**Competing interests** The authors have no competing interests to declare that are relevant to the content of this article.